\documentclass[11pt]{article}
\usepackage[T2A]{fontenc}
\usepackage[cp866]{inputenc}
\usepackage[english]{babel}
\usepackage[dvips]{graphicx }
\usepackage{amsmath,euscript,amssymb}
\newcommand{\be}{\begin{equation}}
\newcommand{\ee}{\end{equation}}
\newcommand{\bea}{\begin{eqnarray}}
\newcommand{\eea}{\end{eqnarray}}
\def\vh{\varphi}

\title{The CMB  Initial
Data  and
 Fundamental Operator Quantization
 of General Relativity}

\author{Victor N. Pervushin,\\
Bogoliubov Laboratory of Theoretical Physics,\\
Joint Institute for Nuclear Research, 141980 Dubna, Russia}

\begin{document}

\maketitle

\begin{abstract}

The observational  data on CMB radiation revealed that our
Universe can be an ordinary physical object moving with respect to
the Earth observer with the occasional initial data. This fact
allows us to apply the theory of irreducible
  unitary representations of the Poincare group in order to
  describe the Universe
   in the framework of the fundamental operator
quantization of General Relatvity and  Standard Model
  of  elementary particles.
 The simplest fit of the observational
CMB data is given that includes occasional gauge-invariant and
frame-covariant initial data and their units of measurement.

\end{abstract}

\vspace{4cm}

\hfil submitted to

\vspace{.51cm}

\hfil {\small XXXIII INTERNATIONAL CONFERENCE ON HIGH ENERGY
PHYSICS}

 \hfil {\small Particle astrophysics \&
cosmology }

\vspace{1cm}

 \hfil {\small ICHEP'06 Moscow, 26.07- 02.08, 2006}

 \hfil  {http://ichep06.jinr.ru}

 \vspace{1cm}

\newpage

\tableofcontents

\section*{Introduction}

  The measurements of the
 Cosmic Microwave Background (CMB) radiation
  temperature  \cite{WMAP} revealed
   the comoving frame of the Universe
   distinguished by its dipole component measured in the rest frame
 of an Earth observer, i.e. his ``Hubble''
  Telescope. We
 believe that the velocity and
 and initial position are occasional values that have no relation
 to the equations of motion and their fundamental parameters of
 the type of the Planck mass.  One can say that the occasional
 motion of the CMB
  comoving frame with respect to the rest frame
  of an observer returned back  us to the historical
 pathway of physics, where occasional  initial data
  are considered as  ingredients supplementing equations of motion
 of any physical theory including the General
 Relativity (GR) \cite{Y,pvng8a,pvng8b}.
  The  occasional  initial data  should not depend on
   the equations of motion (i.e., laws of
 nature) including their fundamental parameters  of
 the type of the Planck mass formed by the Newton coupling constant in GR.

 This historical tradition to separate any theory
 into two independent parts ``equations'' and ``data''
  gives us another admissible alternative
   to the main supposition of the
 Inflationary Model \cite{linde}, according to which the initial data
 of the Universe evolution in GR are
 defined by the  Planck mass as one of the fundamental parameters
 of the equations of motion.

  In  contrast to the  Inflationary Model, the revelation
  of the comoving frame of our Universe allows us to seek
  explanations of  cosmological problems, or part of them,
   with the help of the ordinary Laplace-type questions:
  What are  primordial values
  of the cosmological scale factor (that can be considered here as
  a  gauge-invariant evolution
  parameter \cite{pvng8b}) and its velocity? What are the units of measurement of the
  gauge-invariant and relativistic covariant initial data
  which can give us the simplest fit of all observational data?

 The description of the Universe
  in its comoving frame of reference
   allows us to apply for the construction of gauge-invariant
   variables the theory of irreducible
  unitary representations of the Poincare group and
  {\it fundamental operator quantization} of  GR and Standard Model (SM)
  of  elementary particles
  developed  by Dirac, Heisenberg,
 Pauli, and Schwinger in the case of QED (see \cite{Dir,sch2,pol}).
 A question arises: What does
 {\it fundamental operator quantization} mean
  for both  the SM and GR?
 The answers to all these questions are the topics of the paper.
 We shall list here numerous results on
  the fundamental quantization of GR obtained by the founders
  of  modern theoretical physics and
 give a set of theoretical and observational arguments
 in favor of that this quantization of GR and SM can
 describe the creation of the Universe from
 vacuum in the frame of reference associated with
  the CMB radiation.

\section{Fundamental Operator Quantization}

\subsection{Fundamental Operator Quantization of Electrodynamics and Initial Data}

 Now the point of view is  generally accepted that
 physical results in any gauge theory
 should not depend on any ``frame of reference''.
  This point of view is based on the Faddeev -- Popov heuristic
      quantization \cite{fp1}
      in the frame free Lorentz gauge, as these results
      do not depend on ``gauge''.

      What is wrong in this assertion? Wrong is  slang:
      ``frame of reference''. The right complete phrase is
      ``frame of reference   to initial data'';
    therefore,  physical results as solutions
    of the equations of motion depend on the initial data, given in
    a concrete  frame, and
    the very equations of motion do not depend on the frame, i.e initial
    data, as Galilei defined frame transformations as the ones of
    initial data. The relativistic transformations are only
    the generalization of the Galilei ones.
  Therefore, the problem of initial data
  in any relativistic field theory could  be solved in
  the  comoving frame distinguished by the
  unit  time-like vector ($l^{(0)}_\mu=(1,0,0,0)$).

  In particular, in electrodynamics, this unit  time-like vector
  separates  vector field components $A_\mu$ into time-like $A_0$ and
  the space-like ones $A_j$. The latter are split,
 in agreement with the theory of
 irreducible representation of the Poincare group \cite{ss},
  on two transverse degrees of  freedom $A^{(\rm T)}_i=
  (\delta_{ij}-\partial_i\frac{1}{\triangle}\partial_j)A_j$
  with gauge-invariant initial data (called the ``photon'')
   and the longitudinal part $A^{(||)}_i=
   \partial_i\frac{1}{\triangle}\partial_jA_j$
   which together with the time-like component $A_0$ form
   the gauge-invariant Coulomb  potential $A_0^{(\rm
   T)}=A_0-\partial_0\frac{1}{\triangle}\partial_jA_j
   $ without initial data.  Dirac \cite{Dir} called this gauge-invariant
   functionals $A^{(\rm T)}_\mu,\Psi^{(\rm T)}
   =\exp\{ie\frac{1}{\triangle}\partial_jA_j\}\Psi$ the ``dressed'' fields.
    The propagator of the ``dressed'' vector field
    \cite{Dir,pol}\footnote{The similar relativistic-covariant construction for a massive
    vector field propagator was given in \cite{pvn6}
    \begin{equation}\nonumber
\label{sm} D^R_{\mu\nu}(q) = \delta_{\mu 0}\delta_{\nu 0}{1\over
({q_k}^2+M^2)} +\delta_{\mu i}\delta_{\nu j}
\left(\delta_{ij}-{q_iq_j\over ({q_k}^2+M^2)}\right){1\over
q_0^2-q_k^2-M^2-i\varepsilon}.
\end{equation} }
 \begin{equation}
\label{qed} D^R_{\mu\nu}(q) = \delta_{\mu 0}\delta_{\nu 0}{1\over
{q_k}^2} +\delta_{\mu i}\delta_{\nu j}
\left(\delta_{ij}-{q_iq_j\over {q_k}^2}\right){1\over
q_0^2-q_k^2-i\varepsilon}~.
\end{equation}
  contains the instantaneous Coulomb
 interaction that forms instantaneous atoms and molecules.
 Nobody proved that
   these bound states  formed by the
  Coulomb  potential
  can be obtained by the  Faddeev-Popov heuristic
  quantization  \cite{fp1} in the frame free ``Lorentz gauge
 formulation'' (where all field components are considered
  as ``degrees of freedom'' on equal footing
  and all photon propagators have only the light cone singularities).
 In fact, Faddeev \cite{f1} proved the
 equivalence of this frame free gauge formulation with the Dirac
 {\it fundamental operator quantization} (\ref{qed}) \cite{Dir,pol}
  only for the scattering amplitudes of elementary particles \cite{pvn4,pvn3}.

 To his
   great surprise, a contemporary theoretician may
  know that  Schwinger
 ``{\it  rejected all Lorentz gauge formulations
as unsuited to the role of providing the fundamental operator
quantization}'' \cite{sch2}.

In the context of a consistent description of
   bound states and collective evolution
 of the type of cosmological expansion,
 one can ask what  ``{\it  the fundamental operator
quantization}'' of GR is?

\subsection{GR in the Comoving Frame of Reference}

 Einstein \cite{einsh}
proposed  GR \be\label{gr} S_{\large \rm GR}[\vh_0|F]=\int
d^4x\sqrt{-g}\left[-\dfrac{\vh_0^2}{6}R(g)
 +{\cal L}_{(\rm M)}\right],
 \ee
where $\vh_0=M_{\rm
 Planck}\sqrt{3/{8\pi}}$ is the Planck
mass parameter, as
 generalization of the Lorentz frame group, whereas Hilbert  \cite{H}
 formulated GR so that Einstein's
 generalization $x^\mu\to \widetilde{x}^\mu
 =\widetilde{x}^\mu(x^\mu)$ became a gauge group.
 Recall the principal difference between
 the frame transformations and the gauge ones.
 Parameters of  frame transformations are treated as measurable quantities
 of type of initial data,
  whereas parameters of the gauge  transformations
 (i.e., diffeomorphisms) are not measured.  Diffeomorphisms
  mean that a part of variables are converted into
   potentials without initial data due to  constraints \cite{Noter}.
 All observable initial data should be gauge-invariant.

 The separation of the frame  transformations (here the Lorentz ones)
 from the gauge transformations (here the general coordinate ones)
 can be fulfilled by using  the gauge-invariant components
  of Fock's symplex $\omega_{(\alpha)}$ defined as
$$
 ds^2=g_{\mu\nu}dx^\mu dx^\nu=\omega_{(0)}\omega_{(0)}-
 \omega_{(b)}\omega_{(b)};~~~~~~~\omega_{(\alpha)}=e_{(\alpha)\nu} dx^\nu
 $$
 where
 $e_{(\alpha)\nu}$ are
 the Fock tetrad the components of which   are marked by
 the
 general coordinate index without a bracket and
 the Lorentz index in  brackets $(\alpha)$ \cite{fock29}.

 The choice of   the time axis $l_{(\mu)}=(1,0,0,0)$ as the
 CMB comoving frame allows us to construct
 an irreducible representation of the Poincare group by
  decomposition of Fock's  vector simplex field $\omega_{(\alpha)}$
  in accordance with the  definition of the Dirac Hamiltonian
  approach to GR \cite{dir}
  \be\label{om}
  \omega_{(0)}=\psi^6N_{\rm d}dx^0\equiv \psi^2\omega^{(L)}_{(0)},~~~
  \omega_{(b)}=\psi^2 {\bf e}_{(b)i}(dx^i+N^i dx^0)\equiv\psi^2
  \omega^{(L)}_{(b)},
  \ee
 where $N^i$ is shift vector,
 $N_{\rm d}$ ~~~is Dirac's lapse function, $\psi$ is the spatial metric
 determinant, ${\bf e}_{(b)j}$ is a triad with
 the unit determinant $|{\bf e}|=1$,
 and $\omega^{(L)}_{(0)},\omega^{(L)}_{(b)}$ are
 the scale-invariant Lichnerowicz simplex \cite{Y}
 forming the scale-invariant volume
 $
 dV_0 \equiv
 \omega^{(L)}_{(1)}\wedge \omega^{(L)}_{(2)}\wedge\omega^{(L)}_{(3)}=d^3x
 $
 that coincides with the spatial coordinate volume.
  This classification allowed
  Dirac to separate  all components with zero canonical momentum
 ($N_{\rm d},N_{(b)}=N^j{\bf e}_{(b)j}$) in the GR action   as the potentials
 (i.e. variables without the initial data)
  obtained by the resolution of
 the energy constraint
 \be\label{1ec}
 \frac{\delta S}{\delta N_{\rm d}}\equiv -T_0^0=0
 \ee and momentum one
   \be\label{1mc}
   \frac{\delta S}{\delta N_j}\equiv -T_0^j=0.
   \ee
 Equations
\bea\label{1pc}
 \frac{\delta S}{\delta \log \psi}&\equiv & -T_{\psi}=0,\\
 \label{1ee}
 \frac{\delta S}{\delta {\bf e}_{(b)j}}&\equiv & -T_{(b)}^j=0
 \eea
 can be considered as the dynamic ones.

   \subsection{Gauge-Invariant Variables and Spatial Coordinates}

 It is well known \cite{Y} that the construction of
 tensor and vector variables invariant with respect to
 the general spatial coordinate transformations $x^j \to
 \widetilde{x}^j=\widetilde{x}^j(x^0,x^j)$ repeats the Dirac construction of
 QED with the one-to-one
 correspondence $[ A^{(\rm T)}_0,A^{(\rm T)}_k]~\to~[N^{(\rm T)}_{(b)},
 e^{(\rm T)}_{(b)k}]$.
  Eqs. (\ref{1ee}) describe only
   two transversal gravitons distinguished by the constraint
 \be\label{c1}
\partial_i{\bf e}^{(\rm T)i}_{(b)}\simeq 0,
 \ee
 and constraint (\ref{1mc})  determines only
 two  potentials from three $N^{j}$ (whereas
 the scalar potential $\partial_j [\psi^6N^j]$ can be arbitrary)
 \cite{Y,pvng8a}.
 This means that the  spatial coordinates
 and the Lichnerowicz volume $V_0=\int d^3x$ can be identified
 with  observable in accordance with the Dirac definition \cite{Dir}.

\subsection{Cosmological Scale Factor as a Gauge-Invariant Evolution Parameter}

  In contrast to tensor and vector components,
  the scalar sector $N_{\rm d},\psi,\partial_j [\psi^6N^j]$ in GR
   goes out from the analogy with QED. The problem is the invariance
  of  GR action in the comoving frame
   with respect to the reparametrizations of the
 coordinate evolution parameter: $x^0 \to
 \widetilde{x}^0=\widetilde{x}^0(x^0)$ \cite{6}. This means that
 the coordinate evolution parameter $x^0$ is not
 observable. Therefore, the role of a time-like variable is played by
   one of the dynamic variables.
  A similar situation  is in Special Relativity
 (SR), where  the role of a timelike variable is played by
   one of the dynamic
 variables  $X_0$ in the World space of events $[X_0|X_k]$.
 The reparametrization invariance  leads to the energy constraint
 the
 solution of which with respect to the canonical momentum
 of the timelike variable is identified with the energy
 of the relativistic particle. Recall that the primary and secondary
  quantizations of the energy constraint give   the
   quantum field theory of particle creation as a physical consequence
   of the unitary irreducible representation of the Poincare group.

 Wheeler and DeWitt \cite{WDW1,WDW2} proposed considering the
 reparametrization invariance
 in GR in a similar manner, i.e. they proposed to generalize
 the construction of the unitary irreducible representation
 of the Poincare group in the Minkowskian space of events $[X^0|X^i]$
 to the field space of events in GR. In the case of the  finite space considered in the modern
 cosmology for description of the Universe, the role of
 reparametrization-invariant evolution parameter
   (i.e. a time-like variable in the field space of events) is
   played by
  a  cosmological scale factor $a(x^0)$. This factor is
  separated by the scale transformation of all fields with a conformal
  weight $n$ including
 the metric components
 $$
 F=a^n(x^0) \widetilde{F}^{(n)},~~~~~~
 g_{\mu\nu}=a^2(x^0)\widetilde{g}_{\mu\nu}
 $$
 This transformation keeps the momentum constraint
 $T^k_0=\widetilde{T}^k_0=0$, so that
 the cosmological scale factor $a(x^0)$ can be
 considered as the zero mode solution
 of the momentum constraint \cite{pvn7}.

 The separation of the cosmological scale factor is well-known
 as the ``cosmological perturbation theory''
 (where $\widetilde{\psi}=1-\Psi/2,\widetilde{\psi}^6\widetilde{N}_{\rm d}=1+\Phi$)
  proposed by Lifshits
 \cite{lif} in 1946 and applied now for analysis of
 observational data in modern astrophysics and cosmology
  (see  \cite{bard}). It is generally accepted \cite{lif,bard} that
 the cosmological scale factor is an additional variable without
 any constraint for the deviation $\Psi$, so that
 $\int d^3x \Psi\not = 0$, and there are two zero Fourier harmonics
 of the determinant logarithm
 instead of one. This doubling
 in the ``cosmological perturbation theory''  \cite{bard}
 does not allow to express the velocities of
 both variables $\log a$ and $\int d^3x \Psi$ through
 their momenta and to construct the Hamiltonian approach to GR
 \cite{pvng8a,pvng8b}. In order to restore GR, the logarithm of
 the cosmological scale factor is identified with the Lichnerowicz spatial
 averaging of the  spatial determinant logarithm
 \be\label{csf}
  \log a = \langle \log \psi^2\rangle\equiv \frac{1}{V_0}\int d^3x\log
  \psi^2.
 \ee
 The local part of the spatial determinant logarithm
 $\log\widetilde{\psi}^2\equiv\log \psi^2-\langle \log \psi^2\rangle$
 satisfies the equality
 \be\label{lsf}
 \frac{\delta S}{\delta \log \widetilde{\psi}}\equiv
 -T_{\psi}+\langle T_{\psi}\rangle=0,~~~~~~~~
  \int d^3x\log\widetilde{\psi}^2\equiv 0.
 \ee
 This theory was called in  \cite{pvng8b}
 the ``Hamiltonian cosmological perturbation
 theory''.

 A scale transformation of a curvature
 $
 \sqrt{-g}R(g)=a^2\sqrt{-\widetilde{g}}R(\widetilde{g})-6a
 \partial_0\left[{\partial_0a}\sqrt{-\widetilde{g}}~\widetilde{g}^{00}\right]$
   converts action (\ref{gr}) into
 \be\label{1gr}
 S=\widetilde{S}-
 \int\limits_{V_0} dx^0 (\partial_0\vh)^2\int {d^3x}{\widetilde{N}_{\rm d}}^{-1},
 \ee
 where $\widetilde{S}$
  is the action (\ref{gr})  in
 terms of metrics $\widetilde{g}$ and
 the running scale  of all masses
 $\vh(x^0)=\vh_0a(x^0)$. The variation of this action with respect
 to the new lapse function leads to a new energy constraint
 \be\label{nph}
  T^0_0=\widetilde{T}^0_0-
  \dfrac{(\partial_0\vh)^2}{\widetilde{N}_{\rm d}^2}=0
  ~~~~~~~~
  \left(\widetilde{T}^0_0=-
  \frac{\delta \widetilde{S}}{\delta \widetilde{N}_{\rm d}}\right).
 \ee
 The spatial averaging of the square root
 $\sqrt{\widetilde{T}^0_0}=\pm\dfrac{(\partial_0\vh)}{\widetilde{N}_{\rm d}}$
 over the Lichnerowicz volume $V_0=\int d^3x$
 gives  the Hubble-like relation
 \be\label{Lint}
 \zeta_{(\pm)}=\int dx^0\langle \widetilde{N}_{\rm d}^{-1}\rangle^{-1}=
  \pm\int^{\vh_0}_{\vh}
  d\vh/\langle{(\widetilde{T}_0^0})^{1/2}\rangle,
  \ee
   where
  $\langle F\rangle=V_0^{-1}\int d^3x F$
  and  $d\zeta=\langle(\widetilde{N}_{\rm d})^{-1}\rangle^{-1}dx^0$ is a
time-interval invariant with respect to
  time-coordinate transformations
  $x^0 \to \widetilde{x}^0=\widetilde{x}^0(x^0)$.
  We see that the  Hubble law in the exact
 GR  appears as  spatial averaging of the energy constraint (\ref{nph}).
 Thus, in the contrast to with the generally accepted Lifshits
 theory \cite{lif} its Hamiltonian version \cite{pvng8b} distinguishes
 the variant time-coordinate $x^0$ as an object of reparametrizations
 from the reparametrization-invariant  time interval
 (\ref{Lint})\footnote{Just this interval can be identified
 in GR with the  spatial averaging of
 the conformal time of observable photons as we shall see below.}.

 Just this distinction converts
 the local part of the energy constraint (\ref{nph})
 into equation determining unambiguously
 the gauge-invariant
  Dirac lapse function
 \be\label{13ec}
 N_{\rm inv}={\langle(\widetilde{N}_{\rm d})^{-1} \rangle
 \widetilde{N}_{\rm d}}=
 {{\langle{(\widetilde{T}_0^0})^{1/2}\rangle}}
 (\widetilde{T}_0^0)^{-1/2}.
 \ee

The explicit dependence of $\widetilde{T}_0^0$ on
$\widetilde{\psi}$
  can be given in terms of the scale-invariant  Lichnerowicz
  variables  \cite{Y}
  $\omega^{(L)}_{(\mu)}=\widetilde{\psi}^{-2}\omega_{(\mu)},
  F^{(Ln)}=\widetilde{\psi}^{-n}\widetilde{F}^{(n)}$
 \be\label{t00}
 \widetilde{T}^0_0= \widetilde{\psi}^{7}\hat \triangle \widetilde{\psi}+
  \sum_I \widetilde{\psi}^Ia^{I/2-2}\tau_I, \ee
   where $\hat \triangle
 \widetilde{\psi}\equiv({4\varphi^2}/{3})\partial_{(b)}\partial_{(b)}\widetilde{\psi} $ is
 the Laplace operator and
  $\tau_I$ is partial energy density
  marked by the index $I$ running a set of values
   $I=0$ (stiff), 4 (radiation), 6 (mass), 8 (curvature), 12
   ($\Lambda$-term)
in accordance with a type of matter field contributions, and $a$
is the scale factor \cite{pvng8a}.

 The expression
 $(\widetilde{T}_0^0)^{1/2}$ is Hermitian
 if  a negative
 contribution of  the local determinant momentum
 \bea
\label{gauge}
 p_{\widetilde{\psi}}&=&\frac{\partial {\cal L}}{\partial (\partial_0\log\widetilde{\psi})}\equiv
 -\frac{4\vh^2}{3}\cdot\frac{\partial_l(\widetilde{\psi}^{6}N^l)-
 \partial_0(\widetilde{\psi}^{6})}{\widetilde{\psi}^{6}\widetilde{N_{\rm d}}},
 \eea
is removed  from the energy density (\ref{nph}) by the minimal
surface constraint \cite{Dir}
 \be\label{hg} {p_{\widetilde{\psi}}}\simeq 0 ~~~~\Rightarrow ~~~~
\partial_j(\widetilde{\psi}^6{\cal N}^j)=(\widetilde{\psi}^6)'
~~~~~ ({\cal N}^j=N^j\langle \widetilde{N}_{\rm d}^{-1}\rangle).
 \ee
One can see that the scalar sector $N_{\rm
int},\widetilde{\psi},\partial_j [\widetilde{\psi}^6{\cal N}^j]$
is completely determined in terms of gauge-invariant quantities by
the equations (\ref{lsf}), (\ref{13ec}) and (\ref{hg}), where
${T}_{{\psi}}$ in Eq. (\ref{lsf}) is given as \be\label{tp}
 {T}_{{\psi}}|_{(p_{\widetilde{\psi}}=0)}=7N_{\rm inv}
 \widetilde{\psi}^{7}\hat \triangle \widetilde{\psi}+
  \widetilde{\psi}\hat \triangle[N_{\rm inv}\widetilde{\psi}^{7}]+
  \sum_I I\widetilde{\psi}^Ia^{I/2-2}\tau_I. \ee

These equations are in agreement with the Schwarzschild-type
solution for the potentials
$\triangle\widetilde{\psi}=0,\triangle[N_{\rm
inv}\widetilde{\psi}^{7}]=0$ in the empty space $\tau_I=0$, but
they strongly differ
  from the ``gauge-invariant'' version
\cite{bard} of the Lifshits perturbation
 theory   \cite{lif}.

 First of all, the Lifshits theory \cite{bard}
 considering the scale factor as an additional variable contains
 double counting of the homogeneous variable that is an obstruction to the
 Hamiltonian method and quantization.

 Second, the Lifshits theory \cite{bard} confuses
  the variant time-coordinate with the
 gauge-invariant observable ``conformal time''
 of a cosmic photon. This confusing loses the relativistic
 treatment of the scale factor $\vh$ as the evolution parameter
  in the field space of events
 $[\vh|F^{(L)}]$, where
 its canonic momentum is
  the energy  as the solution of the energy
 constraint. In other words, the Lifshits theory \cite{bard}
 loses the standard Hamiltonian solution of
 the problem of the initial data for the scale factor
 $\vh$  \cite{pvng8a}.

 Third, the Lifshits theory \cite{bard}
 does not take into account
  both the Dirac  constraint (\ref{hg}) removing
  the negative energy of the spatial determinant and
   the potential scalar perturbations formed by the
    determinant
   $\widetilde{\psi}=1+(\mu-\langle\mu\rangle) $ in (\ref{t00}), where
$\sum_I \widetilde{\psi}^Ia^{I/2-2}\tau_I=\sum_n
c_n(\mu-\langle\mu\rangle)^n\tau_{(n)}$,
 $\tau_{(n)}
 \equiv\sum_II^na^{\frac{I}{2}-2}\tau_{I}\equiv \langle\tau_{(n)}\rangle +
 \overline{\tau}_{(n)}$, and
 $\overline{\tau}_{(n)}={\tau}_{(n)}-\langle\tau_{(n)}\rangle$.
 The Hamiltonian cosmological perturbation theory \cite{pvng8a} leads
 to the scalar potentials
\bea\label{12-17}
 \widetilde{\psi}&=&1+\frac{1}{2}\int d^3y\left[D_{(+)}(x,y)
\overline{T}_{(+)}^{(\psi)}(y)+
 D_{(-)}(x,y) \overline{T}^{(\psi)}_{(-)}(y)\right],\\\label{12-18}
 N_{\rm inv}\widetilde{\psi}^7&=&1-\frac{1}{2}\int d^3y\left[D_{(+)}(x,y)
\overline{T}^{(N)}_{(+)}(y)+
 D_{(-)}(x,y) \overline{T}^{(N)}_{(-)}(y)\right],
  \eea
 where $ \beta=\sqrt{1+[\langle \tau_{(2)}\rangle-14\langle
 \tau_{(1)}\rangle]/(98\langle \tau_{(0)}\rangle)}$,
 \be\label{1cur1}\overline{T}^{(\psi)}_{(\pm)}=\overline{\tau}_{(0)}\mp7\beta
  [7\overline{\tau}_{(0)}-\overline{\tau}_{(1)}],
 ~~~~~~~
 \overline{T}^{(N)}_{(\pm)}=[7\overline{\tau}_{(0)}-\overline{\tau}_{(1)}]
 \pm(14\beta)^{-1}\overline{\tau}_{(0)}
 \ee
 are the local currents, $D_{(\pm)}(x,y)$ are the Green functions satisfying
 the equations
 \bea\label{2-19}
 [\pm \hat m^2_{(\pm)}-\hat \triangle
 ]D_{(\pm)}(x,y)=\delta^3(x-y)-\frac{1}{V_0},
 \eea
 where $\hat m^2_{(\pm)}= 14 (\beta\pm 1)\langle \tau_{(0)}\rangle \mp
\langle \tau_{(1)}\rangle$.

  These Hamiltonian solutions (\ref{12-17}) and (\ref{12-18})
  do not contain
the Lifshits-type kinetic scalar perturbations
 explaining the CMB spectrum in the Inflationary Model
 \cite{bard}; they disappear due to the
 positive energy  constraint (\ref{hg}). Therefore, the problem arises
  to reproduce the  CMB spectrum
   by  the fundamental operator quantization.

   In contrast to the Lifshits theory,
    the solutions (\ref{12-17}) and (\ref{12-18}) contain
    the nonzero shift-vector ${\cal N}^i$ of the coordinate origin
   with the spatial metric oscillations that lead  to the new mechanism of formation
  of the large-scale structure of the Universe \cite{pvng8a,pvng8b}.

\subsection{The GR ``Energy'' and Quantum Universe}


One can see that the spatial averaging of the energy constraint
(\ref{nph}) in terms of the scale factor canonical momentum
 \bea \label{pph}
 P_\vh&=&\frac{\partial L}{\partial (\partial_0\vh)}
 = -2V_0\partial_0\vh
 \left\langle(\widetilde{N}_d)^{-1}\right\rangle=
 -2V_0\frac{d\varphi}{d\zeta}\equiv-
2V_0 \vh',
 \eea
takes the form
  $P^2_\vh-E_\vh^2=0$, where  $E_\vh=2\int d^3x(\widetilde{T}_0^0)^{1/2}$.
 Finally, we get the field space of events $[\vh|\widetilde{F}]$,
 where $\vh$ is the evolution parameter, and its canonical momentum
 $P_{\vh}$
 plays the role of the  Einstein-type energy.

  The primary quantization of the energy
  constraint $[\hat P^2_\vh-E_\vh^2]\Psi_L=0$ leads  to
  the unique  wave function $\Psi_L$
  of the
  collective cosmic motion.
 The secondary quantization
 $\Psi_{\rm
L}=\dfrac{1}{\sqrt{2E_\vh}}[A^++A^-]$ describes   creation of a
``number'' of universes
  $<0|A^+A^-|0>=N$
  from the stable Bogoliubov vacuum  $B^-|0>=0$, where $B^-$ is
   Bogoliubov's operator of annihilation of the universe
obtained by the transformation
 $ A^+=\alpha
 B^+\!+\!\beta^*B^-$ in order to diagonalize  equations of
 motion. This causal quantization with the minimal energy
 restricts the motion of the universe in the field space of events
 $E_\vh > 0, \vh_0>\vh_I$ and $E_\vh < 0, \vh_0<\vh_I$, and it
 leads to the arrow of the time interval $\zeta \geq 0$ as the
  quantum anomaly \cite{pvng8a,gip}.

   \subsection{Hamiltonian Reduction and Problem of the Initial Data}
One can construct the  Hamiltonian form of the action (\ref{1gr})
\be\label{hf} S=\int dx^0\left[\int d^3x \left(\sum_F
P_F\!\partial_0
F\!+\!C\!-\!\widetilde{N}_d\widetilde{T}^0_0\right)\!-\!P_{\varphi}\partial_0\varphi+
\frac{P_{\varphi}^2}{4\int dx^3 ({\widetilde{N}_d})^{-1}}\right],
\ee in terms of momenta $P_{ F}=[{p_{\widetilde{\psi}}},
p^i_{{(b)}},p_f]$ and $P_\vh$ given by (\ref{gauge}) and
(\ref{pph}),
  where
 ${\cal C}=N^i {\widetilde{T}}^0_{i} +C_0p_{\widetilde{\psi}}+ C_{(b)}\partial_k{\bf
e}^k_{(b)}$
  is the sum of constraints
  with the Lagrangian multipliers $N^i,C_0,~C_{(b)}$ and the energy--momentum tensor
  components $\widetilde{T}^0_i$; these constraints include
   the transversality  $\partial_i {\bf e}^{i}_{(b)}\simeq 0$ and the Dirac
 minimal  surface \cite{dir} (\ref{hg}).

 One can find
 evolution of all field variables $F(\vh,x^i)$  with respect to
 $\vh$ by the variation of the ``reduced'' action obtained as
   values of the Hamiltonian form of the initial action  (\ref{hf}) onto
 the energy constraint  (\ref{nph}):
 \be\label{2ha2} S|_{P_\vh=\pm E_\vh} =
 \int\limits_{\vh_I}^{\vh_0}d\widetilde{\vh} \left\{\int d^3x
 \left[\sum\limits_{  F}P_{  F}\partial_\vh F
 +\bar{\cal C}\mp2\sqrt{\widetilde{T}_0^0(\widetilde{\vh})}\right]\right\},
\ee
 where $\bar{\cal C}={\cal
 C}/\partial_0\widetilde{\vh}$ and $\vh_0$ is the
 present-day datum that has no relation to the the initial
 data at the beginning $\vh=\vh_I$.
 The reduced action (\ref{2ha2}) shows us
 that the initial data at the beginning $\vh=\vh_I$ are independent of
  the  present-day ones at  $\vh=\vh_0$;
  therefore
  the proposal about the existence of the  Planck epoch $\vh=\vh_0$
   at the beginning \cite{linde} looks
  very doubtful in the framework of the Hamiltonian theory.
  Let us consider consequences of
  the classical reduced theory (\ref{2ha2}) and quantization of
  the energy constraint (\ref{nph}) without the ``Planck epoch''
  at the beginning, proposing occasional data $\vh=\vh_I,\vh'_I$ in
  agreement with the historical tradition and the Hamiltonian theory.

\section{\label{s-4}Observational Data in  Terms of Scale-Invariant Variables}

 Let us assume that the  density
 $T_0^0=\rho_{(0)}(\vh)+T_{\rm f}$
 contains a tremendous  cosmological background
$\rho_{(0)}(\vh)$.
 The  low-energy decomposition
  of ``reduced''  action (\ref{2ha2})  $2 d\vh \sqrt{\widetilde{T}_0^0}= 2d\vh
\sqrt{\rho_{(0)}+T_{\rm f}}
 =
 d\vh
 \left[2\sqrt{\rho_{(0)}}+
 T_{\rm f}/{\sqrt{\rho_{(0)}}}\right]+...$
 over
 field density $T_{\rm f}$ gives the sum
 $S|_{P_\vh=+E_\vh}=S^{(+)}_{\rm cosmic}+S^{(+)}_{\rm
 field}+\ldots$, where the first  term of this sum
 $S^{(+)}_{\rm cosmic}= +
 2V_0\int\limits_{\vh_I}^{\vh_0}\!
 d\vh\!\sqrt{\rho_{(0)}(\vh)}$ is  the reduced  cosmological
 action,
 whereas the second one is
  the standard field action of GR and SM
 \be\label{12h5} S^{(+)}_{\rm field}=
 \int\limits_{\zeta_I}^{\zeta_0} d\zeta\int d^3x
 \left[\sum\limits_{ F}P_{ F}\partial_\eta F
 +\bar{{\cal C}}-T_{\rm f} \right]
 \ee
 in the  space determined by the  interval
 $
 ds^2=d\zeta^2-\sum_a[e_{(a)i}(dx^i+{\cal N}^id\zeta)]^2;
 ~\partial_ie^i_{(a)}=0,~\partial_i{\cal N}^i=0
  $
 with  conformal time
 $d\eta=d\zeta=d\vh/\rho_{(0)}^{1/2}$ as the gauge-invariant
 and scale-invariant quantity, coordinate distance
 $r=|x|$,
 and running masses
 $m(\zeta)=a(\zeta)m_0$.
 We see that
  the  correspondence principle leads to the theory,
  where the scale-invariant conformal  variables and coordinates are
    identified  with the observable ones
    and the cosmic evolution with the evolution of masses:
  $$
\frac{E_{\rm emission}}{E_0}=\frac{m_{\rm atom}(\eta_0-r)}{m_{\rm
atom}(\eta_0)}=\frac{\vh(\eta_0-r)}{\vh_0}=a(\eta_0-r)
=\frac{1}{1+z}.
$$
The conformal observable distance  $r$ loses the factor $a$, in
comparison with the nonconformal one $R=ar$. Therefore, in this
 case, the redshift --
  coordinate-distance relation $d\eta=d\vh/\sqrt{\rho_0(\vh)}$
  corresponds to a different
  equation
  of state than in the standard one  \cite{039,Danilo}.
    The best fit to the data  including
  cosmological SN observations \cite{SN2,SN1}
 requires a cosmological constant $\Omega_{\Lambda}=0.7$,
$\Omega_{\rm CDM}=0.3$ in the case of the Friedmann
``scale-variant quantities`` of standard cosmology, whereas for
the ``scale-invariant conformal
 quantities''
 these data are consistent with  the dominance of the stiff state
of a free scalar field $\Omega_{\rm Stiff}=0.85\pm 0.15$,
$\Omega_{\rm CDM}=0.15\pm 0.10$ \cite{039}. If $\Omega_{\rm
Stiff}=1$, we have the square root dependence of the scale factor
on conformal time $a(\eta)=\sqrt{1+2H_0(\eta-\eta_0)}$. Just this
time dependence of the scale factor on
 the measurable time (here -- conformal one) is used for a description of
 the primordial nucleosynthesis \cite{Danilo,three}.
Thus, the stiff state formed by a free scalar field
\cite{039,Danilo} can describe  in the relative (conformal) units
all epochs including the creation of a quantum universe at
$\vh(\eta=0)=\vh_I,H(\eta=0)=H_I$.

\section{Creation of Matter  and Initial Data of the Universe}
 The initial data $\vh_I,H_I$ of the universe can be  determined
 from the
 parameters of matter cosmologically created from the stable
 quantum
 vacuum  at the beginning of the universe.

 1. The Standard
 Model in the framework of the perturbation theory
 and the fundamental operator quantization of SM
  \cite{pvn6} shows us  that  W-,Z-vector bosons
 have maximal probability of the
 cosmological creation due to their mass singularity.
 The uncertainty principle $\triangle E\cdot\triangle \eta \geq 1$
  (where $\triangle E=2M_{\rm I},\triangle \eta=1/(2H_{\rm I})$)
 testifies that these bosons can be created
 from vacuum at   the moment when   their  Compton   length
 defined by the inverse mass
 $M^{-1}_{\rm I}=(a_{\rm I} M_{\rm W})^{-1}$ is close to the
 universe horizon defined in the
 stiff state as
 $H_{\rm I}^{-1}=a^2_{\rm I} (H_{0})^{-1}$.
 Equating these quantities $M_{\rm I}=H_{\rm I}$
 one can estimate the initial data of the scale factor
 $a_{\rm I}^2=(H_0/M_{\rm W})^{2/3}=10^{-29}$ and the Hubble parameter
 $H_{\rm I}=10^{29}H_0\sim 1~{\rm mm}^{-1}\sim 3 K$ \cite{pvng7,pvng8}.


2. The collisions and scattering processes with the cross-section
 $\sigma \sim 1/M_{\rm I}^2$ can lead to conformal
 temperature $T_c$. This temperature
   can be estimated  from the condition that  the relaxation time
 is close to the life-time of the universe, i.e.,
 can be estimated from the equation in the
 kinetic theory $\eta^{-1}_{relaxation}\sim n(T_c)\times \sigma \sim H $.
 As
 the distribution functions of the longitudinal   vector bosons
 demonstrate a large contribution of relativistic momenta
 \cite{pvng8} $n(T_c)\sim T_c^3$,
 this kinetic equation  gives the temperature of
relativistic bosons $
 T_c\sim (M_{\rm I}^2H_{\rm I})^{1/3}=(M_0^2H_0)^{1/3}\sim 3 K
$ as a conserved number of cosmic evolution compatible with the SN
data \cite{039}.
 We can see that
this  value is surprisingly close to the observed temperature of
the CMB radiation
 $ T_c=T_{\rm CMB}= 2.73~{\rm K}$.
 The equations describing the longitudinal vector bosons
 in SM, in this case, are close to
 the equations that  are  used in  the
 Inflationary Model \cite{bard} for
 description of the ``power primordial spectrum'' of the CMB radiation.

 3. The primordial mesons before
 their decays polarize the Dirac fermion vacuum and give the
 baryon asymmetry frozen by the CP -- violation
 so that $n_b/n_\gamma \sim X_{CP} \sim 10^{-9}$,
 $\Omega_b \sim \alpha_{\rm \tiny QED}/\sin^2\theta_{\rm Weinberg}\sim
 0.03$, and $\Omega_R\sim 10^{-4}$~\cite{pvng8}.

 All these results
 testify to that all  visible matter can be a product of
 decays of primordial bosons, and the observational data on CMB
  reflect rather  parameters of the primordial bosons, than the
 matter at the time of recombination. In particular,
 the length of  the semi-circle on the surface of  the last emission of
photons at the life-time
  of W-bosons
  in terms of the length of an emitter
 (i.e.
 $M^{-1}_W(\eta_L)=(\alpha_W/2)^{1/3}(T_c)^{-1}$) is
 $\pi \cdot 2/\alpha_W$.
 It is close to the value of
 orbital momentum with the maximal $\Delta T$:
 $l_{(\Delta T_{\rm max})}\sim \pi \cdot 2/\alpha_W\sim  210 $,
 whereas $(\bigtriangleup T/T)$ is proportional to the inverse number of
emitters~
 $(\alpha_W)^3 \sim    10^{-5}$.

The temperature history of the expanding universe looks
 in relative units like the
 history of evolution of masses of elementary particles in the cold
 universe with the constant conformal temperature $T_c=a(\eta)T=2.73~ {\rm K}$
 of the cosmic microwave background.

 The  nonzero shift vector and
  the scalar potentials given by Eqs. (\ref{12-17}) and (\ref{12-18})
    determine
 in relative units \cite{039} the parameter
  of spatial oscillations
  $ m^2_{(-)}=\frac{6}{7}H_0^2[\Omega_{\rm R}(z+1)^2+\frac{9}{2}\Omega_{\rm
  Mass}(z+1)]$. The redshifts in the recombination
  epoch $z_r\sim 1100$ and the clustering parameter
 $
 r_{\rm clust.}={\pi}/{ m_{(-)} }\sim {\pi}/[{
 H_0\Omega_R^{1/2} (1+z_r)}] \sim 130\, {\rm Mpc}
 $
  recently
 discovered in the researches of large scale periodicity in redshift
 distribution \cite{da}
 lead to a reasonable value of the radiation-type density
  (including the relativistic baryon matter one)
  $10^{-4}<\Omega_R\sim 3\cdot 10^{-3}$ at the time of this
  epoch.

\section*{Conclusion}

The
 observational astrophysical data on CMB radiation
  revealed that our Universe can be an ordinary physical
  object moving with respect to the Earth observer
   with occasional initial data.
  This revelation returns  us back to the historical
  traditions of physics beginning with
     Ptolemaeus' rest frame, Copernicus' comoving
  one, and Galilei's frame transformations as the initial data ones,
  and finishing by
 representations of the Poincare group as the basis of
 {\it fundamental operator quantization} that
  includes occasional gauge-invariant and
  frame-covariant initial data and their units of measurement.
 In the context of the history of physics as a whole, in order
 to explain the World,
 a modern Laplace should ask for
{\it  the initial data of the gauge-invariant
 variables measured  in the
 relative units in the comoving frame of reference of this World}.

 The gauge-invariant
 variables mean here nothing but the application of the standard theory
 of the unitary irreducible representations of the Poincare group
 based on the time-like unit vector that distinguishes the
 comoving frame in the Minkowskian space-time.
 Another frame means a choice of another time-like unit vector connected with the
 first one by the Lorentz transformation that leads to the dipole
 component of the CMB temperature \cite{WMAP}.

  We
 gave here  a set of numerous arguments in favor of that
 the fundamental operator quantization
  can be a real theoretical basis
 for a   further detailed investigation of astrophysical
 observational data, including
  CMB fluctuations.


{\bf Acknowledgements}\\
 The author is grateful to B. Barbashov, A. Zakharov, and V. Zinchuk for
 collaboration,
 and thanks  V. Kadyshevsky, A.~Efremov,
 V.~Priezzhev,   I.~Tkachev, and S. Vinitsky
 for fruitful discussions.

\end{document}